# Acoustic-Driven Magnetic Skyrmion Motion

Yang Yang[1#], Le Zhao[2#], Di Yi[3#], Teng Xu[2], Yahong Chai[1], Chenye Zhang[1], Dingsong Jiang[1], Yahui Ji[1], Wanjun Jiang[2*], Jianshi Tang[1], Pu Yu[2], Huaqiang Wu[1], Tianxiang Nan[1*]

**Magnetic skyrmions have great potential for developing novel spintronic devices. The electrical manipulation of skyrmions has mainly relied on current-induced spin-orbit torques. A recent theoretical model suggested that the skyrmions could be more efficiently manipulated by surface acoustic waves (SAW), an elastic wave that can couple with magnetic moment through magnetoelastic effect. However, the directional motion of skyrmions that is driven by SAW is still missing. Here, we experimentally demonstrate the motion of Néel-type skyrmions in Ta/CoFeB/MgO/Ta multilayers driven by propagating SAW pulses from on-chip piezoelectric transducers. Our results reveal that the elastic wave with longitudinal and shear vertical displacements (Rayleigh wave) traps skyrmions, while the shear horizontal wave effectively drives the motion of skyrmions. In particular, a longitudinal motion along the SAW propagation direction and a transverse motion due to topological charge, are observed and further confirmed by our micromagnetic simulations. This work demonstrates a promising approach based on acoustic waves for manipulating skyrmions, which could offer new opportunities for ultra-low power spintronics.**

## Main

Using magnetic skyrmions, the particle-like spin textures, as controllable information carriers offer potentials for high density and low power spintronic memory and logic applications[1-8]. To develop skyrmion-based devices, *e.g.* skyrmion racetrack memory, efficient manipulation of skyrmions is crucial. Electric current manipulation of skyrmions has been previously demonstrated in asymmetric magnetic multilayers by means of current-induced spin-orbit torques or thermal gradients[9-20]. On the other hand, the electric-field control of magnetization via magnetoelectric or magnetoelastic effect could provide more energy-efficient approaches for manipulating the spin texture with an extremely low Joule heating and hence low power consumption[21,22]. Such control means can be achieved, for example, by the static strain modification of magnetic anisotropy and Dzyaloshinskii–Moriya (DM) interaction[23-25], or by dynamic strains using acoustic waves through strong magnon-phonon coupling[26,27]. In particular, surface acoustic waves (SAWs) are long-range carriers (wave propagating over millimeter distances through ferromagnets) for dynamic strains[28-36], which have been used as an efficient source for generating skyrmions by the SAW-induced spatiotemporally varying strains and inhomogeneous effective torques[37]. A recent theoretical model also suggested the skyrmion motion driven by counter-propagating SAWs[38]. Yet the electric-field induced static strain or acoustic wave control of the skyrmion motion has not been realized. Though SAWs have been used as a noncontact and controllable method to manipulate nano/microparticles[39,40], electrons[41], and qubits[42], the associated efficiency seems not high enough to induce the motion of magnetic skyrmions.

Here, we demonstrate the SAW-driven motions of Néel-type magnetic skyrmions due to the strong magnetoelastic coupling in magnetic multilayers integrated with on-chip piezoelectric transducers. By controlling the relative orientation of acoustic wave propagation and crystal orientation of the piezoelectric materials, we generate both Rayleigh waves (with both shear vertical and longitudinal displacements) and shear horizontal (SH) waves (with only shear horizontal displacements) that can be applied to skyrmions at the same sample area. We find the Rayleigh wave can generate but not move the skyrmions due to its dominant vertical displacement (Fig. 1a), consistent with the previous report[37]. By contrast, the SH wave can efficiently move skyrmions as a result of the strong magnetoelastic coupling induced by the in-plane strain gradients. The observed directional motion shows a longitudinal component along the wave propagation direction, and a transverse component with the sign depending on the topological charges, in analogy to the skyrmion Hall effect[12,43]. These experimental observations are further confirmed by our micromagnetic simulations. Our results not only provide an efficient approach to drive the skyrmion motion by electric field-induced strain wave, but also demonstrates the SAW could serve as a versatile platform to explore the skyrmion dynamics.

The difference of the skyrmion motion driven by Rayleigh and SH waves can be captured by the micromagnetic simulations by considering magnetoelastic coupling, exchange coupling and DM interaction (see Method and Supplementary Information note 2 for detail). Figs. 1b and 1e show the simulated spatial distribution of the out-of-plane normalized magnetization component $m_z$, magnetoelastic energy density and total energy density for skyrmions that were driven by Rayleigh and SH waves, respectively. We find that the Rayleigh wave can only move skyrmions for less than half an acoustic wavelength (the skyrmions would then be trapped at the antinodes of the Rayleigh wave). The amplitude of the shear vertical displacement is usually much larger than that of the longitudinal displacement in a Rayleigh wave, in which the shear vertical displacement traps the skyrmions. The magnetoelastic energy density distribution of a skyrmion under a shear vertical wave shows a left-right asymmetry that is different from that under an SH wave including both left-right and up-down asymmetry due to the different displacement modes. The total energy density of the skyrmion (including the magnetoelastic energy density, anisotropy energy density, magnetostatic energy density, exchange and DM energy density) illustrates a symmetric distribution under a shear vertical wave, while an asymmetric distribution along the diagonal axis under an SH wave which causes the skyrmion to move towards the lower energy density direction.

[1] School of Integrated Circuits and Beijing National Research Center for Information Science and Technology (BNRist), Tsinghua University, Beijing, China.
[2] Department of Physics, Tsinghua University, Beijing, China.
[3] School of Materials Science and Engineering, Tsinghua University, Beijing, China.
# These authors contributed equally.
* Corresponding author, e-mail: jiang_lab@tsinghua.edu.cn; nantianxiang@mail.tsinghua.edu.cn

We study the skyrmions driven by SAWs in Ta (5 nm)/$Co_{20}Fe_{60}B_{20}$ (CoFeB, 1 nm)/MgO (1 nm)/Ta (2 nm) multilayers, because the multilayers show a weaker pinning effect than that in Pt/Co/Ta multilayers[10,20], and the amorphous CoFeB has a strong magnetoelastic coupling and a low damping parameter, simultaneously[44]. The magnetic properties of the multilayers on a $LiNbO_3$ substrate were characterized by a polar magneto-optic Kerr effect (MOKE) magnetometry (Supplementary Fig. S1a). The average diameter of skyrmions generated by magnetic field pulses in the sample is estimated to be around 1 μm (Supplementary Fig. S1c). The multilayers and interdigital transducers (IDTs) were integrated on a 64°Y-cut $LiNbO_3$ piezoelectric substrate, as shown in Fig. 1h (optical image of the fabricated devices is shown in Supplementary Fig. S1d). By controlling the angle between the SAW propagation direction and the orientation of the piezoelectric substrate, the piezoelectric constant matrix can be transformed (see Supplementary Information Note 2), and thus the SH wave or Rayleigh wave can be generated independently with the wave propagation along the *x* or *y* direction, respectively (Fig. 1d). Fig. 1i shows a transmission spectrum ($S_{21}$) between two IDTs using a vector network analyzer where the resonance frequencies of the SH wave mode and the Rayleigh wave mode are 486 MHz and 451 MHz, respectively.

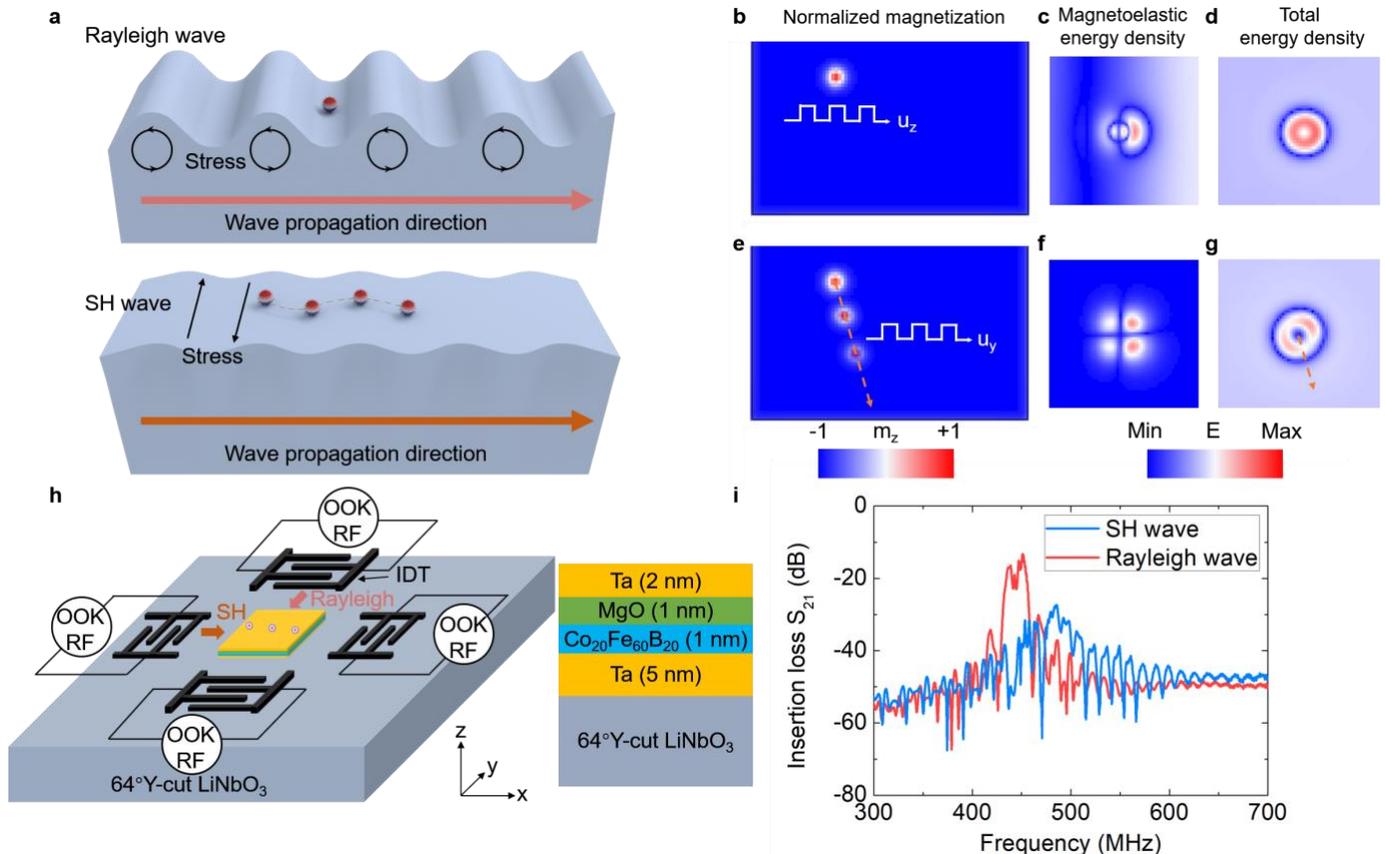

**Fig. 1 | Concept of skyrmion motion driven by SAWs and the device schematics. a**, Schematic diagram of skyrmion motion driven by a Rayleigh wave or a shear horizontal (SH) wave. **b**, Simulated skyrmion pinning under an elastic wave with periodic shear vertical displacements. The diameter of the simulated skyrmion is set to be 30 nm. The wavelength is 240 nm which is eight times as large as the simulated skyrmion size. The color scale represents the normalized (out-of-plane) magnetization component $m_z$. **c,d** The magnetoelastic energy density (**c**) and the total energy density (**d**) of the skyrmion under an elastic wave with periodic shear vertical displacements. **e**, Simulated skyrmion motion driven by an SH wave. The arrow denotes the direction of the skyrmion motion trajectory. **f,g** The magnetoelastic energy density (**f**) and the total energy density (**g**) of the skyrmion under an SH wave. **h**, Schematic of the SAW delay lines configuration and the structure of the magnetic multilayer. **i**, Transmission spectra ($S_{21}$) of the SH and Rayleigh wave. The wavelength of SAWs is 8 μm.

## Skyrmion generation by Rayleigh and SH waves

We first study skyrmion generation by using SAWs. Fig. 2 shows the MOKE images for the evolution of topological magnetic textures. At a zero magnetic field, magnetic textures are the maze domains (Fig. 2a). Then we start with a state with almost no magnetic texture by eliminating the initial maze domain structure (by the applied out-of-plane magnetic field of -0.8 mT), as shown in Fig. 2b. Magnetic skyrmions with a topological charge Q=+1 are created after exciting a propagating Rayleigh wave or SH wave with a pulse duration of 300 ms at the resonance frequencies, as shown in Figs. 2c and 2d. The positive or negative sign of Q represents the center magnetization of the skyrmion being up or down. Figs. 2e and 2f show the estimated skyrmion densities and sizes created by SH and Rayleigh waves as a function of applied RF powers. Note that the skyrmion density created by a Rayleigh wave is slightly smaller than that in the previous report[37], because our pulse duration is smaller. We find that the SH wave can generate skyrmions much more efficiently than the Rayleigh wave when RF power is above 20 dBm. This indicates that the SH wave mode can couple with skyrmions more effectively because the SH wave with its dominant in-plane shear horizontal displacement produces stronger in-plane magnetoelastic energy. The average sizes of skyrmions generated by both SH and Rayleigh waves are estimated to be around 1 μm, similar to that generated by the magnetic field.

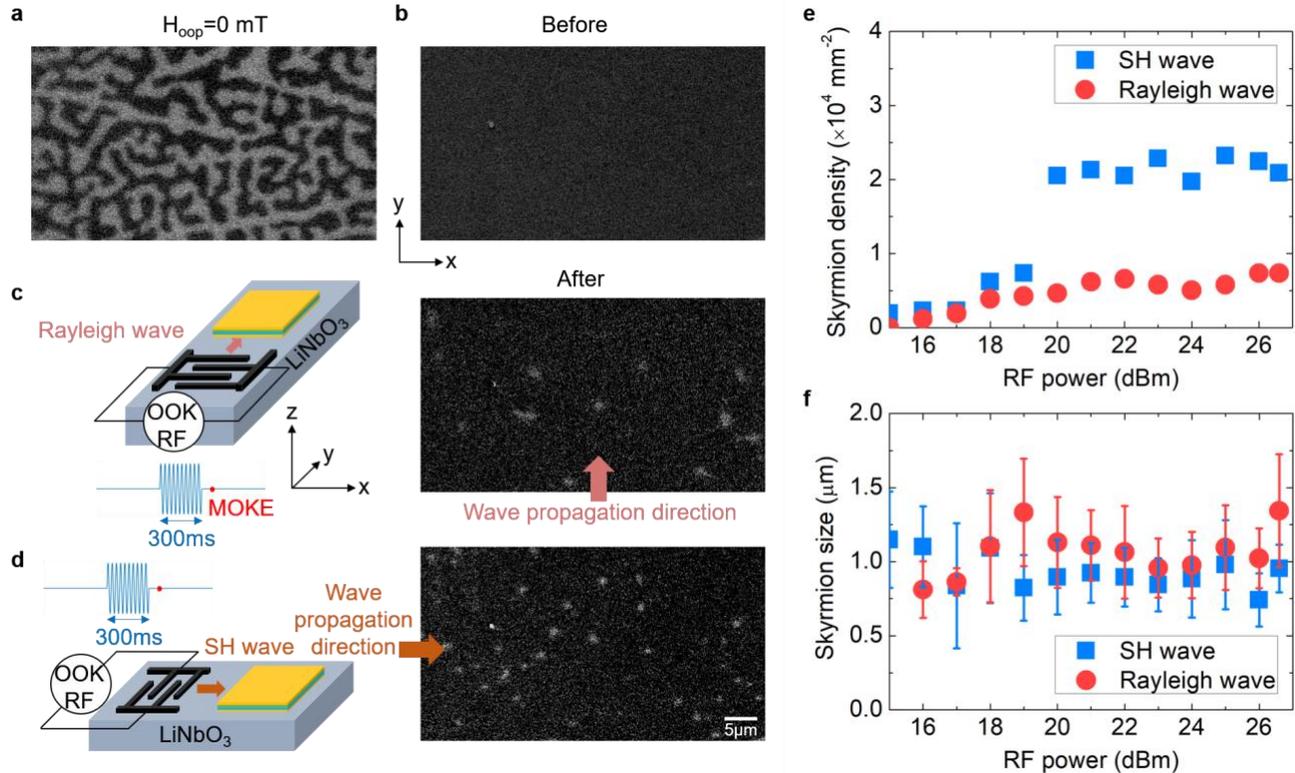

**Fig. 2 | Generation of skyrmions by Rayleigh and SH waves. a-d**, Polar MOKE images of the maze domain at a zero magnetic field (**a**), before exciting any SAW wave with an applied out-of-plane magnetic field $H_{oop}$=-0.8 mT (**b**), after exciting a propagating Rayleigh wave (**c**), and after exciting a propagating SH wave (**d**). The RF power is fixed at 26 dBm. The SAW pulse duration is 300 ms. The scale bar is 5 μm. **e,f** The density (**e**) and the size (**f**) of skyrmions created by SH waves and Rayleigh waves as a function of RF powers.

## Skyrmion motion by SH waves

We then study the skyrmion motion by applying continuous SAW pulses with a fixed RF power of 26 dBm. Figs. 3a-d (e-h) illustrate MOKE images of Q=+1 (Q=-1) skyrmion after exciting 1st-4th SH wave pulse with a duration of 300 ms (also see Supplementary movies 1 and 2). We find that the skyrmions move along the wave propagation direction (*x* axis) also with a transverse component (*y* axis), in analogy to the skyrmion Hall effect[12]. The motion distances $d$ ( $d = \sqrt{dx^2 + dy^2}$ ) of the circled skyrmions (Q = ±1) are around 3 μm after each pulse. The SH wave with a wavelength of 10 μm also moves skyrmions (Supplementary Fig. S3). Statistically, 32% of the skyrmion population in the whole sample shows motion distance $d$>1 μm (Supplementary Fig. S4). The motion distances can be further improved by increasing RF powers or decreasing the wavelength of SAWs (Supplementary Fig. S4). In contrast, we do not observe any skyrmion motion driven by Rayleigh waves although the power of the receiving IDTs for a Rayleigh wave is 12 dBm higher than that of an SH wave (Supplementary Fig. S5), which is consistent with our micromagnetic simulations.

We estimate the skyrmion velocity driven by SH wave to be about 10 μm/s, which is similar to that driven by current-induced spin-orbit torques with small current densities[43]. This could suggest that the skyrmion motion driven by SH waves under the current experimental condition remains in the creep regime. By progressively increasing the RF power, it is possible to increase the skyrmion velocity, but the wave amplitude is saturated (Supplementary Fig. S5). To transform from the creep regime to the flow regime with a much higher skyrmion velocity, one can use magnetic films with stronger magnetoelastic coupling constants and low damping parameters (Supplementary Information Note 3).

We summarize the motion trajectories (both $d_x$ and $d_y$) of 18 different magnetic skyrmions in Fig. 4a. The skyrmions with Q=+1 (Q=-1) move consistently along the wave propagation direction with $d_y$<0 ($d_y$>0). This is in agreement with our micromagnetic simulation (Fig. 4b and 4c). The average deflection angles ( $\phi_{sk} = \arctan(d_y/d_x)$ ) of Q=-1 and Q=+1 skyrmions are around 49.5°±15.2° and -34.2°±17.7°, respectively. The large variation of deflection angles is because the motion is influenced by the pinning potential induced by random defects. Our analytical calculation by solving the Thiele equation[46] reveals that the deflection angle is determined by the damping parameter and the ratio of effective magnetoelastic forces along the y and x axes ($F_x/F_y$). In Fig. 4d, the solid curves show the calculated deflection angle as a function of damping parameters with $F_y/F_x = 0.3$ (typical value for an SH wave). The calculated curves correspond to the experimental data (extract from Fig. 4a), which gives the damping parameters in the range of 0.01-0.07 (the calculated deflection angles with different $F_y/F_x$ are shown in Supplementary Fig. S6). The large deflection angles we observed in the creep regime are in sharp contrast with that observed in the current driven experiments (skyrmion Hall effect)[12], as the skyrmion Hall angles are generally very small in the creep regime. This behavior indicates the magnetoelastic effective field can be an additional factor besides the topological Magnus force to determine the deflection angle.

We also find that decreasing the SH wave pulse duration causes the skyrmions to deform. When the duration of the SH wave pulse is reduced to 200 ms, some circular skyrmions deform into strips (see Supplementary Fig. S7). For longer SH wave pulses, more skyrmions are created and the average distance among skyrmions is reduce, causing a stronger skyrmion–skyrmion repulsion and a limited skyrmion motion (Supplementary Fig. S7).

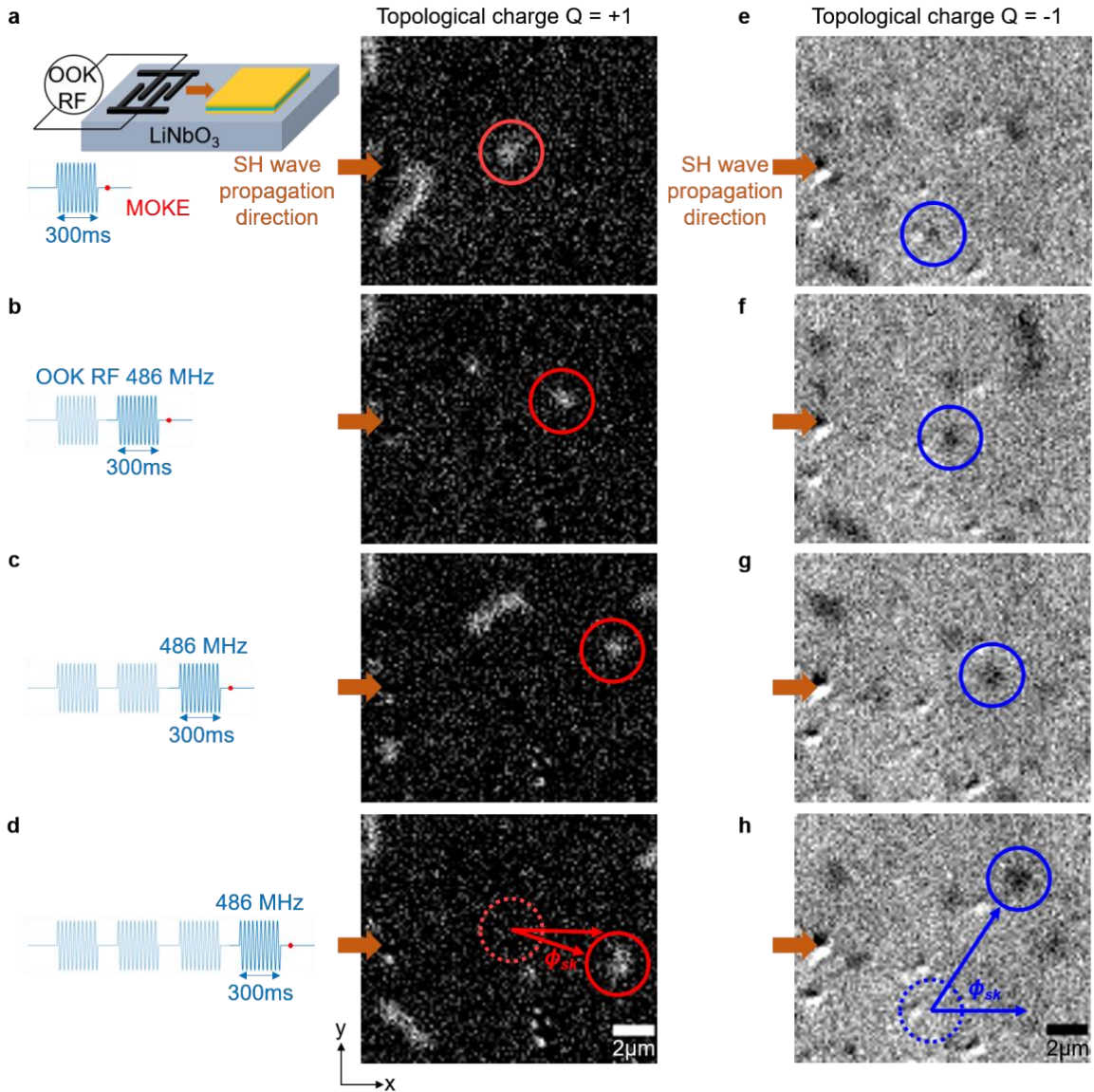

**Fig. 3 | Skyrmion motions driven by SH waves. a-h**, The MOKE images of skyrmions motion for topological charge Q=+1 (**a-d**) and Q=-1 (**e-h**) skyrmions after the exciting of 1st-4th SH wave pulses. The red and blue circles highlight the position of the Q=+1 and Q=-1 skyrmions. The dotted circle in **d** and **h** represent the initial position of the skyrmions shown in **a** and **e**. The RF power is fixed at 26 dBm. The SAW pulse duration is 300 ms. The wavelength of SAWs is 8 μm. The SH wave propagating direction is from left to right. The SAW creates skyrmions with a topological charge Q=±1 when the positive or negative out-of-plane magnetic field ($H_{oop} = ±0.8$ mT) is applied. The scale bar in the MOKE images is 2 μm.

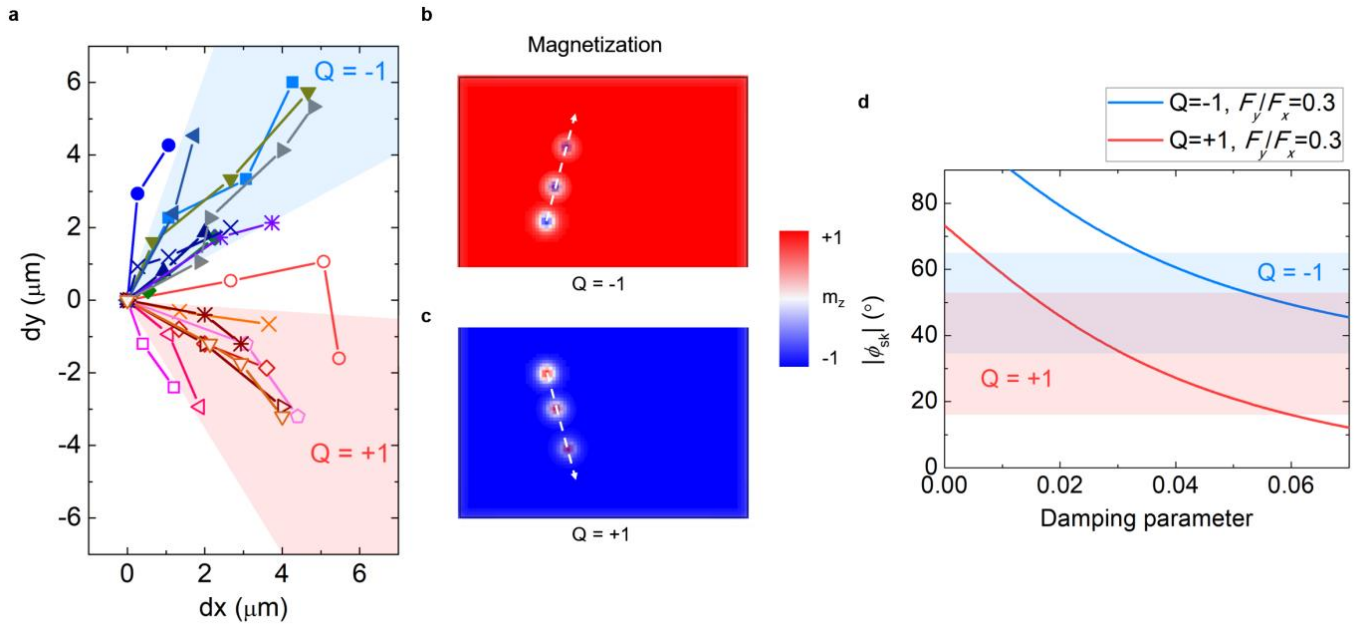

**Fig. 4 | Skyrmion motion trajectories and skyrmion deflection angles driven by SH waves. a**, Motion trajectories of magnetic skyrmions driven by SH waves in experiments. **b,c** Simulated skyrmions motion driven by the SH wave for Q=-1 skyrmions (**b**) and Q=+1 skyrmions (**c**). **d**, The experimental (blue and red regions) and numerically calculated (curves) skyrmion deflection angles ($\phi_{sk}$) versus the damping parameter with a fixed ratio of effective magnetoelastic forces along $y$ and $x$ axes $F_y/F_x$=0.3.

## Conclusions

In summary, we have experimentally demonstrated the motion of Néel-type skyrmions driven by acoustic waves in magnetic multilayers, in which the skyrmions can move along the wave propagation direction with a deflection angle with respect to the wave propagation direction, consistent with our simulations. This work provides insights into the understanding of magnetoelastic or magnetoelectric coupling in skyrmions. This approach of manipulating skyrmion dynamics by acoustic wave will potentially lead to skyrmion-based memory, logic and microwave devices with low power, without any current flowing in magnetic layers, in large sample areas, and with arbitrary motion trajectory (circular motion, Supplementary Fig. S8). The comparison of our work and other skyrmion-driven methods is listed in Supplementary Table. S1. The efficiency of acoustic wave driven skyrmion motion can be further enhanced by materials (with high magnetoelastic coupling) and device (with high power handling) engineering. More importantly, the motion trajectory with controlled directivity and high precision can be achieved possibly on a single skyrmion limit by using high-frequency acoustic waves (small wavelength) and techniques of acoustic wave manipulation, such as phased array acoustic transducers[47]. This is comparable to acoustic tweezers for dynamic microparticle manipulation[48].

## Methods

**Sample fabrication**. Synchronous two-ports SAW delay line devices were patterned on a 64°Y-cut LiNbO$_3$ substrate by using photolithography and a lift-off fabrication process. Ti (5 nm)/Pt (150 nm) electrodes were deposited on a 64°Y-cut LiNbO$_3$ substrate by using a high vacuum magnetron sputtering with Ar pressure of 3 mTorr. The SH type leaky SAW is confined on the surface using IDTs consisting of heavy metal Pt electrodes on top of the LiNbO$_3$ substrate. The magnetic multilayers Ta (5 nm)/Co$_{20}$Fe$_{60}$B$_{20}$ (1 nm)/MgO (1 nm)/Ta (2 nm) were sputtered by using high vacuum magnetron sputtering with an Ar pressure of 3 mTorr.

**P-MOKE measurements with in-situ RF voltages**. The skyrmions were imaged by using a polar magneto-optic Kerr effect (p-MOKE) microscope, as shown in Supplementary Fig. S9. All measurements were performed at room temperature. The transmission spectrum between two IDTs was measured using a vector network analyzer (Keysight E5080B). Radiofrequency (RF) voltage pulses supplied to IDTs were provided by an analog signal generator (Keysight N5183B) and a function/arbitrary waveform generator (Keysight 33210A). The frequency of the RF voltage is the same as the resonance frequency of the SAW. SAW pulses are generated by an on-off keyed (OOK) RF modulation.

**Micromagnetic simulations**. The micromagnetic simulation is implemented including interfacial Dzyalosinskii-Moriya interaction, exchange interaction, magnetic anisotropy, magnetostatic, and magnetoelastic coupling contributions using MuMax3[49-51]. Due to the limitation of the computational capacity, the length of the simulated layer is set as 256 nm, the width is 256 nm, and the height is 1 nm. The discretization cell size along the $x$, $y$, and $z$ axes are 1 nm, 1 nm, and 0.5 nm, respectively. The material parameters used in the simulations are as follows: exchange parameter $A_{ex}$ = 1.8×10$^{-11}$ J/m, saturation magnetization $M_s$ = 5.8×10$^5$ A/m, interfacial Dzyaloshinskii-Moriya interaction strength D = 3×10$^{-3}$ J/m$^2$, perpendicular anisotropy parameter $K_u$ = 6×10$^5$ J/m$^3$, Gilbert damping parameter $\alpha$ = 0.2, a mass density of 8000 kg/m$^3$, the first order and the second order magnetoelastic coupling constants $B_1$ = $B_2$ = -8.8 ×10$^6$ J/m$^3$, elastic constants $C_{11}$ = 283 GPa, $C_{12}$ = 166 GPa, and $C_{44}$ = 58 GPa[44]. The simulations are calculated under ideal conditions without considering elastic wave attenuation. However, the wave attenuation and the pinning effect due to the impurities and defects induced disorder in actual devices lead to skyrmion velocities driven by SAWs will be lower in reality.

## Data availability

The data that support the finding of this study are available from the corresponding author upon reasonable request.

## Acknowledgements

This work was supported by the National Key R&D Program of China (Grant No. 2021YFA0716500), National Natural Science Foundation (Grant No. 52073158, 52161135103, 62131017), and the Beijing Advanced Innovation Center for Future Chip (ICFC).